# RHEED Studies of Epitaxial Oxide Seed-Layer Growth on RABiTS Ni(001): The Role of Surface Structure and Chemistry

C. Cantoni, D. K. Christen, R. Feenstra, A. Goyal, G. W. Ownby, and D. M. Zehner

*Oak Ridge National Laboratory, Oak Ridge, TN 37831*

D. P. Norton

*University of Florida, Department of Materials Science and Engineering, Gainesville, FL 32611*

## Abstract

The epitaxial deposition of the first oxide buffer layer (seed layer) on biaxially textured Ni tape for coated conductors is a critical step that is dependent on the atomistic surface condition of the metal. We present a study of the {100}<100> biaxially textured Ni (001) surface and seed-layer growth using *in situ* reflection high-energy electron diffraction (RHEED) and Auger electron spectroscopy (AES). Our observations are consistent with formation of a c(2 × 2) 2-D superstructure due to surface segregation of sulfur contained in the metal. We show that this superstructure can have a dramatic effect on the heteroepitaxial growth of oxide seed layers. In particular, the surface superstructure promotes the (200) epitaxial oxide growth of $Y_2O_3$-stabilized $ZrO_2$ (YSZ), which is necessary for the development of high-$J_c$ superconducting films for coated conductors.

Fabrication of coated conductors for real applications relies on the industrial scale-up of flexible metal-like substrates suitable for epitaxial growth of high temperature superconducting (HTS) high-$J_c$ films. In the rolling-assisted biaxially textured substrate (RABiTS) approach, such substrates are prepared by deposition of epitaxial oxide buffer layers on a {100}<100> biaxially textured, 99.99% pure Ni substrate[1,2]. Deposition of the first oxide buffer layer (seed layer) on textured Ni is the most delicate step in RABiTS fabrication because the crystalline properties of this layer control those of subsequently deposited layers. Nucleation of an oxide film on a metal surface is a critical and not fully understood process that involves a complex interplay between the thermodynamic stability and kinetic formation of an ionic/non-ionic interface. For example, recent work has shown that YSZ (100) can be nucleated on Ni (100) only under specific vacuum conditions[3].

Although several different seed layer oxides have been successfully grown on textured Ni over the past few years by different techniques[2], the previous works were based on an incomplete understanding of the Ni surface properties. In particular, the seed layer was assumed to nucleate on a clean and pure Ni surface that was obtained by a reducing heat treatment (using, for example, a mixture of Ar and $H_2$) before deposition[1,2,4]. A better understanding of the structure and chemistry of the textured {100}<100> Ni surface, and the effect on seed layer nucleation is essential for reproducibility and scalability of RABiTS.

In this letter, we present for the first time a study of the {100}<100> Ni surface and YSZ seed layer growth using *in situ* reflection high-energy electron diffraction (RHEED) and Auger electron spectroscopy (AES). This study was conducted in two separate ultrahigh vacuum (UHV) chambers. A laser ablation molecular beam epitaxy (MBE) system equipped with RHEED, a mass spectrometer, and a pulsed KrF eximer laser ($\lambda = 248$ nm), was used to deposit the seed layer and to monitor surface structural changes during Ni substrate annealing and oxide nucleation. A second UHV chamber equipped with AES and a sputter-ion gun was used for surface chemical analysis of the {100}<100> Ni substrate at various temperatures. We show that the structure and chemical composition of the textured Ni surface has a profound effect on the heteroepitaxial growth of the oxide seed layer.

After cold rolling and recrystallization annealing[5], cube textured Ni substrates have typical grain sizes and surface roughness of 50−100 μm and 5−10 nm, respectively. The degree of grain alignment is expressed by a full width half maximum (FWHM) of 6−10° for the in-plane and out-of-plane grain boundary misorientation distributions. Previous inductively coupled plasma analyses on similar samples of Ni tape indicated the presence of C and S as major impurities in the bulk, with concentrations less than 100 and 30 wt. ppm, respectively. The samples considered in this study (more than 30) were fabricated in different batches with some variation in impurity content expected.

The as-formed Ni substrates with typical dimensions of $10 \times 20$ mm$^2$ were mounted on a radiant heater plate in air and loaded in the laser MBE chamber through a load lock. At room temperature, with a chamber background pressure of $5-10 \times 10^{-10}$ Torr, no RHEED pattern was observed. This suggests the presence of an amorphous layer on the sample's surface prior to heating. Heating the sample to 500° C in a background pressure of $10^{-8}-10^{-9}$ Torr was sufficient to completely remove the weakly bound adsorbates, resulting in a clear and distinct RHEED pattern. A typical RHEED pattern of the {100}<100> Ni surface with the incident electron beam along the <100> crystal direction is shown in Fig. 1(a). Although the diffraction pattern is broadened due to the orientation distribution of the grains (the electron beam samples a few hundred grains), the RHEED streaks are well-defined and distinct. Scanning the beam over the sample's surface did not change their position. The pattern in Fig. 1(a) shows two extra streaks (indicated by arrows) in addition to the reflections observed for a clean Ni (001) surface pattern. The extra streaks are positioned halfway between the Ni streaks and are not observed when the incident beam is parallel to the <110> direction. This situation is consistent with the presence of a c(2 × 2) superstructure on the {100}<100> Ni surface. Such a superstructure was observed on all the samples examined and could not be removed by annealing in vacuum, or by annealing at 800° C in a mixture of 4% $H_2$ in Ar (up to a maximum investigated total pressure of 200 mTorr). In some cases, the intensity of the extra streaks decreased slightly after annealing in Ar/$H_2$ at high temperature for several hours, but did not disappear completely. A superstructure-free Ni RHEED pattern was obtained by depositing a Ni overlayer *in situ* by pulsed laser deposition (PLD) from a 99.99% pure Ni target. The RHEED pattern for such a surface is shown in Fig. 1(b).

Over the past few decades, a large number of surface studies conducted on single crystal Ni have demonstrated that clean, low-index planes of Ni are not reconstructed. However, C, CO, and many elements of the VI group, such as O and S, chemisorb on the Ni (001) face, forming either p(2 × 2) and c(2 × 2) structures, whereby the adsorbed atoms reside in a plane about 1 Å above the outer Ni plane and are bound in the fourfold hollow[6-10]. Figure 2 shows a schematic of a c(2 × 2) surface structure formed by specie A chemisorbed on Ni (001). An analogous structure can form by diffusion and surface segregation of bulk impurities[11].

To study the effect of such a superstructure on seed layer epitaxy, we deposited YSZ on the c(2×2)/{100}<100> Ni surface, and on a superstructure-free Ni overlayer that was ablated *in situ* on the biaxially textured Ni substrate. Figure 3 shows the X-ray *θ-2θ* patterns and relative pole figures acquired from the YSZ films grown on textured Ni with and without superstructure, respectively. The YSZ films were grown under the same conditions and with the same procedure, while monitoring the process with RHEED. The deposition temperature was 800° C. After an initial ~100-Å-thick layer was deposited in vacuum ($P_{base}=5 \times 10^{-8}$ Torr), the $O_2$ partial pressure was increased to the value $1 \times 10^{-5}$ Torr and a final 1200-Å-thick film was grown. The YSZ films grown on the c(2 × 2) surface showed single (002) orientation, with a (111) pole figure indicating the same degree of grain alignment as the substrate. The resulting YSZ unit cell was rotated 45° in plane with respect to the Ni cell [Fig. 3(a)]. In contrast, the YSZ films grown on the superstructure-free Ni overlayer showed only the (111) peak in the $\theta$-$2\theta$ scan [see Fig. 3(b)]. In this case, the pole figure of the (200) reflection showed 4 different in-plane domains rotated 30° with respect to each other. This epitaxial relation is expected for the nucleation of a threefold symmetric lattice on a square symmetric lattice. The quality of the YSZ seed layer on c(2 × 2)/Ni was tested by growing a 0.3-μm-thick YBCO film by the *ex situ* $BaF_2$ method[12] on some of these samples. A 20-nm-thick $CeO_2$ cap layer enabled compatibility of the precursor layer with the YSZ. The resulting YBCO critical current density was 1.15 MA/cm$^2$ in self-field at 77 K, indicating that a 120-nm-thick YSZ film is a good Ni diffusion barrier and a good buffer layer for coated conductors.

To determine the chemical nature of the observed superstructure, we complemented the RHEED observations with systematic AES analyses both before and after anneals at different temperatures and after ion bombardment to clean the surface of adsorbates. An initial AES spectrum of the samples showed typical O, C, and S peaks in addition to Ni. Since NiO is thermodynamically stable at room temperature in atmospheric oxygen pressure, its formation in the amount of a few monolayers is expected under air exposure. The O peak present in the AES spectra acquired at room temperature disappeared completely once the samples were heated in vacuum to ≥ 500° C, a temperature at which the superstructure was clearly observed by RHEED. This observation excludes the possibility that chemisorbed O or a Ni sub-oxide is responsible for the observed 2-D superlattice. In fact, at these temperatures the AES results showed only S and Ni peaks. The sulfur peak remained after 1-hour anneals in an $H_2$ partial pressure of $10^{-4}$ Torr at 500° C. On the other hand, S was eliminated by sufficient ion bombardment of the substrate surface. This suggests that S was present (in detectable amounts) only in a very thin, superficial layer. Note that a small carbon signal was always detected after ion bombardment, even after repeated treatments, indicating the presence of this contaminant element throughout the bulk. Although the spectra acquired after ion bombardment showed only Ni and C peaks, the S signal was progressively recovered after anneals in vacuum at a temperature of 430° C or higher, and in particular at 800° C where deposition of YSZ takes place. The observation of S after annealing was associated with the decrease and ultimate disappearance of the C peak. Collectively, these observations suggest that the c(2 × 2) superstructure that promotes the YSZ (200) growth is formed by S atoms originating from substitutional sites in the Ni lattice. One may surmise that, while the bulk S is relatively immobile below 400° C, S atoms readily diffuse and segregate to the free surface during the initial Ni fabrication (e.g., during the high-temperature recrystallization anneal)[13, 14]. In contrast, C atoms, which occupy Ni-lattice interstitial positions, are very mobile and segregate to the surface at low temperatures ($T \leq 400°$ C). The surface C adatoms are displaced on formation of the S superstructure, leading to an observed depletion of surface C[14]. The AES analysis of the samples with a Ni overlayer detected segregation of only C after anneals in the temperature range 350–500° C. Sulfur was detected after these samples were annealed for 1 hour in vacuum at temperatures higher than 600° C. This observation suggests that the ablated Ni films contain less S and is consistent with the superstructure-free Ni pattern obtained from the {100}<100> textured Ni samples with the Ni overlayer. In these samples, the segregation of S to the surface (and consequent C depletion) takes place only after the S atoms have diffused through the entire Ni overlayer. Moreover, the arrangement of diffused S atoms into a c(2 × 2) surface structure was deduced by a RHEED experiment in which the c(2 × 2) extra reflections were observed to slowly reform from a previously clean pattern at $T = 750°$ C over a time span of 2 hours.

Figure 4 shows a proposed possible explanation of the c(2 × 2) mediated epitaxial growth of YSZ on the (001) Ni surface. The fluorite structure of YSZ is characterized by the existence of atomic subplanes that contain only oxygen. This square oxygen sublattice matches well with the square net formed by the superstructure adatoms when they occupy every fourfold Ni site. We propose that during deposition, the unoccupied Ni hollows of the c(2 × 2)-S structure are filled with O, and the nucleation of YSZ takes place starting from the S+O overlayer and continues in the sequence (Y+Zr)-plane, O-plane, and so on.

On the basis of this study, it is likely that other epitaxial oxide seed layers ($CeO_2$, $Y_2O_3$) grown on biaxially textured Ni involve nucleation on (100) Ni surfaces characterized by a stable c(2 × 2) sulfur superstructure. Such a superstructure cannot be removed by simply annealing the Ni substrate in forming gas, as is typically done prior to deposition. At the same time, this superstructure is highly beneficial for YSZ seed layer nucleation in that it promotes the (200) epitaxy of this oxide. Remaining important issues for coated conductor development include the reproducibility and control of the superstructure, its importance for the epitaxial deposition of other common seed layers, and its generality with respect to other deposition methods (including solution-based approaches). Research to address these issues is ongoing.

**Figure Captions**

**Fig. 1.** RHEED patterns obtained with the incident electron beam along <100> for: a), RABITS Ni; and b), Ni overlayer deposited by PLD on RABiTS Ni. The arrows in a) indicate the existence of a c(2 × 2) superstructure that is absent in b).

**Fig. 2.** Schematic of the atomic Ni (001) surface with a c(2 × 2) superstructure formed by specie A. The black dots symbolize Ni atoms and the gray dots symbolize A atoms. The black and the gray squares indicate the Ni and the superstructure unit cells, respectively. When the incident electron beam is directed along the <100>, the spacing between the streaks resulting from the c(2 × 2) 2-D lattice is half the distance between the Ni streaks. When the electron beam is directed along the <110>, the streaks originated from the Ni and the c(2 × 2) lattices overlap.

**Fig. 3.** X-ray *θ-2θ* scan and pole figure for: a), the YSZ film grown on the c(2 × 2) superstructure present on RABiTS Ni; and b), the YSZ film grown on the Ni overlayer epitaxially deposited on RABiTS Ni.

**Fig. 4.** Proposed c(2 × 2) mediated epitaxial growth of YSZ on (001) Ni surface. The superstructure adatoms become constituents of the basal oxygen sublattice of YSZ.

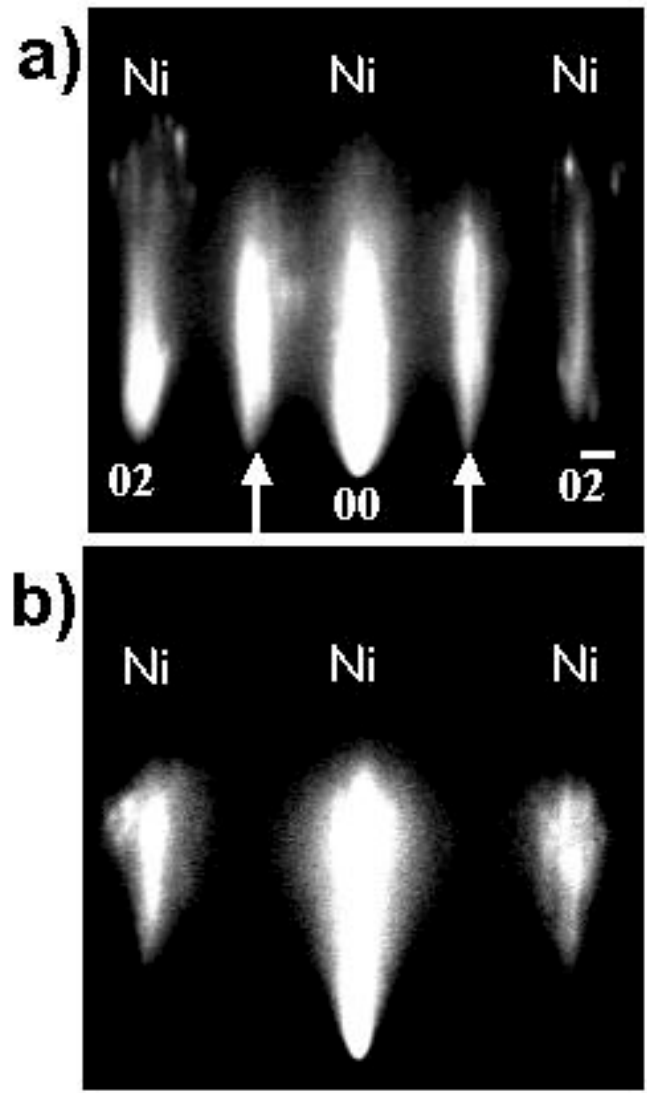

**Fig. 1**

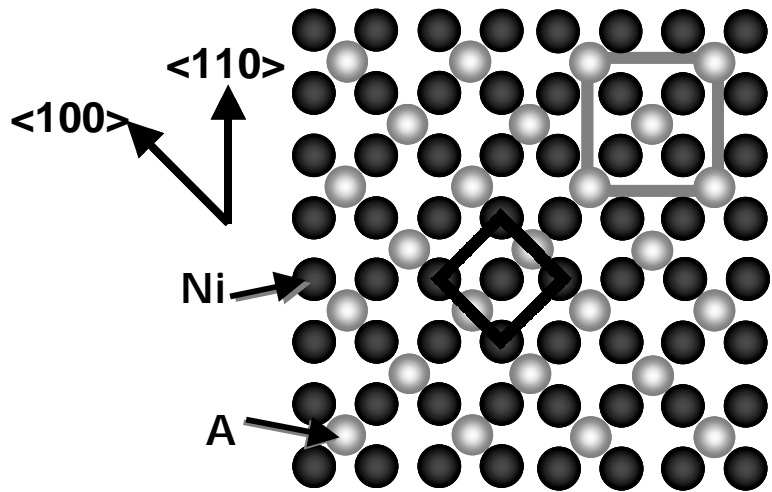

**Fig. 2**

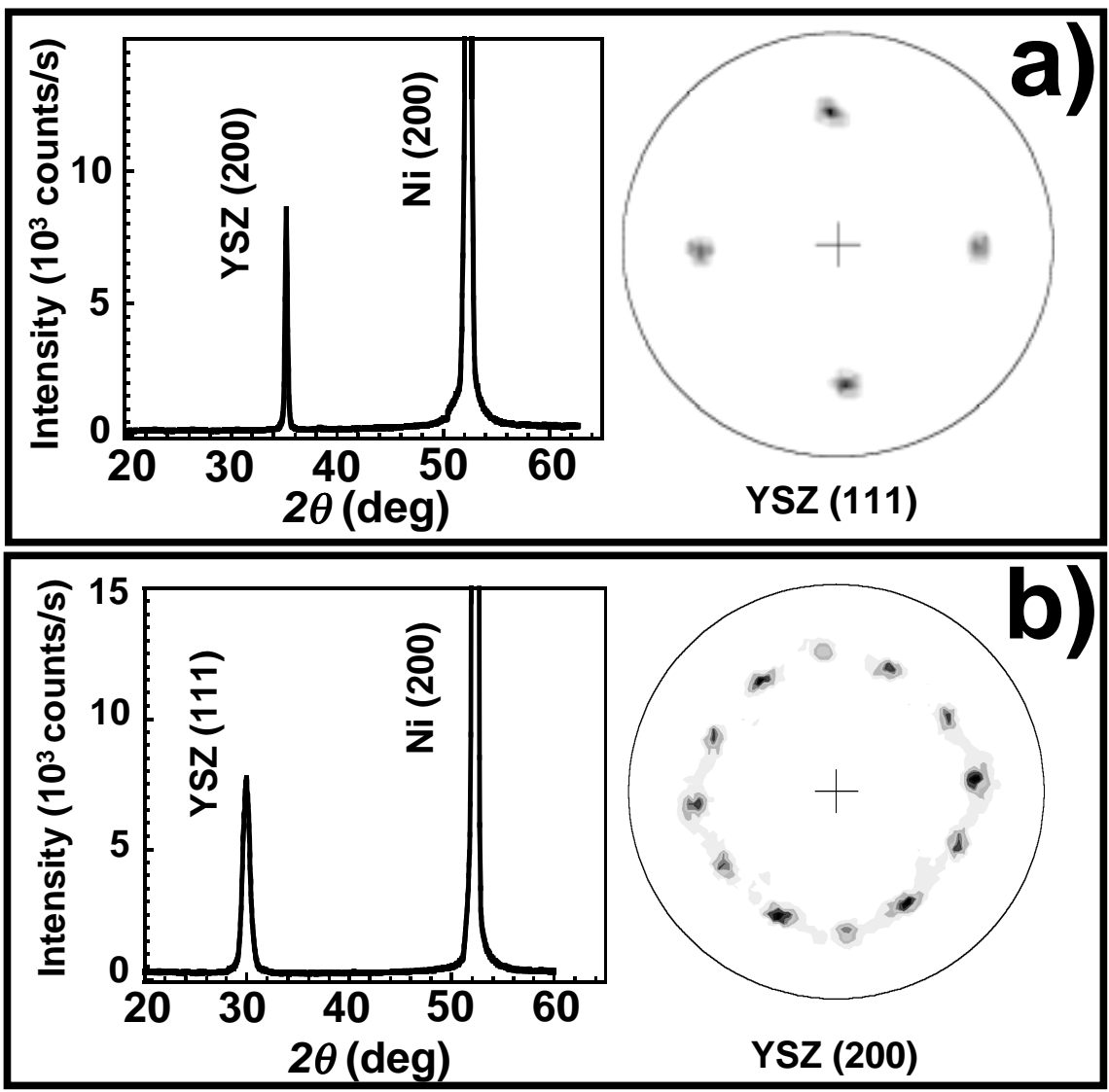

**Fig. 3**

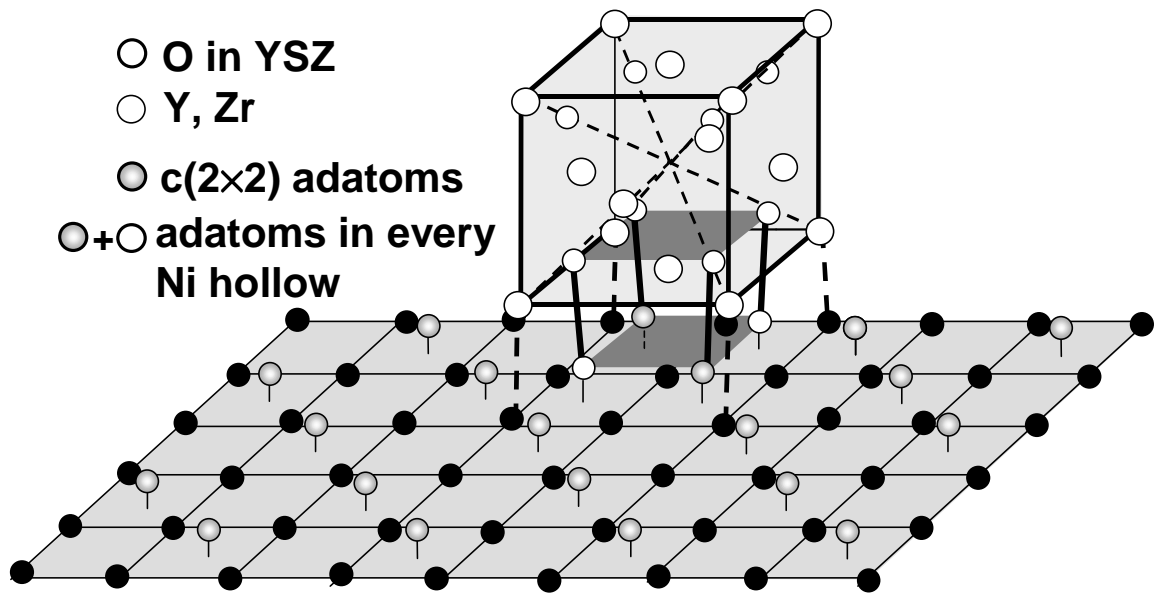

**Fig. 4**